\documentclass{article}
\usepackage{graphicx}%
\usepackage{multirow}%
\usepackage{amsmath,amssymb,amsfonts}%
\usepackage{amsthm}%
\usepackage{mathrsfs}%
\usepackage[title]{appendix}%
\usepackage{xcolor}%
\usepackage{textcomp}%
\usepackage{booktabs}%
\usepackage{algorithm}%
\usepackage{algorithmicx}%
\usepackage{algpseudocode}%
\usepackage{listings}%
\usepackage{enumitem}  
\usepackage{float}
\usepackage{hyperref}
\begin{document}

\title{TockyPrep: Data Preprocessing Methods for Flow Cytometric Fluorescent Timer Analysis}
\author{Masahiro Ono \\ {\footnotesize \texttt{m.ono@imperial.ac.uk}} \\ {\footnotesize Department of Life Sciences} \\ {\footnotesize Imperial College London} \\ {\footnotesize Imperial College Road, London, SW7 2AZ, United Kingdom}}

\maketitle

\begin{abstract}

\textbf{Background:}\\
Fluorescent Timer proteins, which display time-dependent changes in their emission spectra, are invaluable for analyzing the temporal dynamics of cellular events at the single-cell level. We previously developed the Timer-of-cell-kinetics-and-activity (Tocky) tools, utilizing a specific Timer protein, Fast-FT, to monitor temporal changes in cellular activities. Despite their potential, the analysis of Timer fluorescence in flow cytometry is frequently compromised by variability in instrument settings and the absence of standardized preprocessing methods. The development and implementation of effective data preprocessing methods remain to be achieved.

\textbf{Results:}\\
In this study, we introduce the R package that automates the data preprocessing of Timer fluorescence data from flow cytometry experiments for quantitative analysis at single-cell level. Our aim is to standardize Timer data analysis to enhance reproducibility and accuracy across different experimental setups. The package includes a trigonometric transformation method to elucidate the dynamics of Fluorescent Timer proteins. We have identified the normalization of immature and mature Timer fluorescence data as essential for robust analysis, clarifying how this normalization affects the analysis of Timer maturation. These preprocessing methods are all encapsulated within the \texttt{TockyPrep} package.

\textbf{Conclusions:}\\
\texttt{TockyPrep} is available for distribution via GitHub at \url{https://github.com/MonoTockyLab/TockyPrep}, providing tools for data preprocessing and basic visualization of Timer fluorescence data. This toolkit is expected to enhance the utility of experimental systems utilizing Fluorescent Timer proteins, including the Tocky tools.

\end{abstract}

\section{Background}

Fluorescent Timer proteins are distinguished by their unique ability to exhibit time-dependent changes in their emission spectra. A variety of these proteins have been developed, including DsRed-Timer \cite{Terskikh2000}, mCherry-derived monomeric variants such as Fast-FT, Medium-FT, and Slow-FT \cite{Subach2009}, and mRuby variants, known as mRubyFT \cite{Subach2022}. These proteins have enabled detailed analyses of temporal dynamics in molecular and cellular events. Applications range from examining intracellular and membranous protein dynamics \cite{Subach2009} to monitoring T cell and B cell activities in vivo \cite{Bending2018JCB}, and assessing promoter activities \cite{Terskikh2000, Troscher2019}. Furthermore, tandem fluorescent protein timers, pairing two fluorescent proteins with distinct expression and maturation kinetics, are employed in both cellular and in vivo studies. The combination of Superfolder GFP (sfGFP), a GFP variant known for its rapid fluorescence maturation, with a red fluorescent protein exhibiting more delayed fluorescence kinetics, such as mCherry or TagRFP, is a popular choice for these tandem Timer systems \cite{Khmelinskii2012, Barry2015}. The analysis of data from these studies typically involves sophisticated mathematical modeling to elucidate temporal changes in gene expression by exploiting the differing kinetics of the two fluorescent proteins \cite{Barry2015}.

Using the monomeric Fluorescent Timer protein Fast-FT, we have developed the Timer-of-cell-kinetics-and-activity (Tocky) system for single-cell flow cytometric analysis of cellular activities and transcription \cite{Bending2018JCB}. This approach has resulted in the development of Fluorescent Timer reporter mouse strains: Foxp3-Tocky for analyzing the temporal dynamics of Foxp3 transcription and Nr4a3-Tocky for studying cellular responses over time following T cell receptor (TCR) signaling. These developments have established the proof-of-concept for the Tocky systems as valuable experimental methods\cite{Bending2018JCB}\cite{Bending2018EMBO}. However, the full implementation of Timer fluorescence data analysis has yet to be realized.

Currently, the primary bottleneck in the application of Fluorescent Timer proteins lies in data analysis, particularly the flow cytometric analysis of Timer fluorescence, which presents both significant challenges and opportunities for advancement. Fast-FT, an mCherry mutant with delayed protein maturation kinetics, spontaneously and irreversibly transitions its chromophore from blue to a mCherry-type red form \cite{Subach2009}. Analysing the ratio of immature to mature fluorescence provides insights into time-dependent cellular dynamics. We have previously described the trigonometric transformation of Timer fluorescence into the angle component, Timer Angle, and the norm, Timer Intensity, which decomposes temporal information and signal strength\cite{Bending2018JCB}. Nevertheless, the data preprocessing process requires further clarification and standardization. 

Importantly, flow cytometric measurements of fluorescence rely on signal amplification \cite{Snow2004}, and the instrumental settings, especially the voltages used for amplification, can be significantly varied between users \cite{McKinnon2018}. This variability complicates the consistent application of trigonometric transformations, underscoring the necessity for robust and practical data preprocessing methods. Consequently, the normalization of immature and mature Timer fluorescence data becomes crucial in ensuring that the analysis of Timer blue and red is reliable and reproducible across different experimental setups.

In this paper, we introduce a new R implementation of the Timer data analysis method. This implementation automates the data preprocessing methods, including Timer fluorescence normalization and trigonometric transformation, which are essential for analyzing Timer data from flow cytometry experiments. Additionally, tailored visualization techniques are included to ensure accurate data preprocessing. Collectively, these enhancements aim to standardize the process, enhancing both reproducibility and accuracy.

\section{Materials and Methods}

\subsection{Overview of Timer Data Analysis}

Timer data preprocessing consists of three main steps:

\begin{enumerate}
    \item Timer fluorescence thresholding, which determines Timer fluorescence positivity
    \item Timer fluorescence normalization, which normalizes Timer blue and red fluorescence data.
    \item Trigonometric data transformation of blue and red fluorescence into Timer Angle and Timer Intensity using polar coordinates.
\end{enumerate}

The preprocessed data, specifically Timer Angle and Timer Intensity, are subsequently used for downstream analysis.

\subsection{Timer Thresholding}
Setting thresholds for blue and red fluorescence to gate cells:
\begin{itemize}
    \item Threshold values are determined either interactively or through automatic quantile-based methods.
    \item Cells are selected for further analysis based on these thresholds to enable quantitative analysis of Timer dynamics using Timer Angle and Timer Intensity.
\end{itemize}

\subsection{Timer Fluorescence Normalization Using Negative Control Data}

Timer blue and red fluorescence data are normalized using statistics derived from the gated negative control data. Gating thresholds \( x_{\text{lim.red}} \) and \( y_{\text{lim.blue}} \) for red and blue fluorescence are set either interactively or automatically based on quantiles. The maximum values and Median Absolute Deviations (MADs) or standard deviations (SDs) of the log-transformed fluorescence in the gated negative control data are computed:

Using the parameters defined below,  the log-transformed fluorescence values are normalized for each cell:

\begin{equation}
    B_{\text{norm}} = \frac{B_{\text{log}} - \max(B_{\text{log, gated\_neg}})}{\text{SD}(B_{\text{log, gated\_neg}})},
\end{equation}
where $\max(B_{\text{log, gated\_neg}})$ is the maximum value and $\text{SD}(B_{\text{log, gated\_neg}})$ is the SD of the log-transformed blue fluorescence of gated negative cells.

\begin{equation}
    R_{\text{norm}} = \frac{R_{\text{log}} - \max(R_{\text{log, gated\_neg}})}{\text{SD}(R_{\text{log, gated\_neg}})}.
\end{equation}
where $\max(R_{\text{log, gated\_neg}})$ is the maximum value and $\text{SD}(R_{\text{log, gated\_neg}})$ is the SD of the log-transformed red fluorescence of gated negative cells.

\subsection{Timer Fluorescence Trigonometric Transformation}

The core of our temporal analysis is the trigonometric transformation of the normalized fluorescence data into polar coordinates. This step captures the progression of the fluorescent protein's maturation from blue to red over time. The Timer Intensity \( I \) represents the overall expression level and is calculated as:

\begin{equation}
    I = \sqrt{B_{\text{norm}}^2 + R_{\text{norm}}^2}.
\end{equation}

The Timer Angle \( \theta \), representing the temporal aspect, is computed using the arccosine function:

\begin{equation}
    \theta = \arccos\left( \frac{B_{\text{norm}}}{I} \right) \times \left( \frac{180}{\pi} \right).
\end{equation}

This angle effectively maps the maturation state of the fluorescent protein, with values ranging from \( 0^\circ \) (pure blue fluorescence) to \( 90^\circ \) (pure red fluorescence), reflecting the time elapsed since expression.

\subsection{Nr4a3 Tocky Dataset}
In this study, we assessed the effectiveness of Timer fluorescence normalization using flow cytometric data from antigen-stimulated T cells, specifically obtained from Nr4a3-Tocky::OT-II double transgenic mice. These T cells, expressing the OT-II transgenic TCR, were activated using their specific agonist peptide, Ova\textsubscript{(323-339)}
 at a concentration of 1~\(\mu\)M in the presence of antigen presenting cells. This setting provides continuous TCR signals to T cells and induce Fluorescent Timer expression. Time course analysis was performed. This dataset was previously detailed in our earlier publication \cite{Bending2018JCB}, where we demonstrated the initial application of the Tocky system in a biological context. 

\subsection{Computational Performance}
The function \texttt{timer\_transform} was executed on a system equipped with Apple M2 Ultra, 128 GB memory, and MacOS Sonoma 14.7.1. Performance was measured by the `system.time()` function in R, which reported an average user time of 2.278 seconds, system time of 0.266 seconds, and elapsed time of 2.781 seconds. 

\subsection{Software}
The plots in Figures~\ref{fig:fig7} and \ref{fig:fig9} were generated using the R packages \texttt{ggplot2}\cite{ggplot2}, \texttt{gridExtra}\cite{gridExtra}, and \texttt{dplyr}\cite{dplyr}. The package \texttt{RColorBrewer}\cite{RColorBrewer} was used to generate heat colors. All other figures were generated by functions implemented in the \texttt{TockyPrep} package. The interactive application for exploring Timer data preprocessing parameters was developed using the \texttt{Shiny} framework in R \cite{shiny2025}. 

\newpage

\section{Implementation}
\subsection{Fluorescent Timer Protein}

Fluorescent Timer proteins have two types of fluorescence, and the different kinetics of these fluorescence levels are exploited to decipher temporal dynamics of biological systems. We use an mCherry mutant Fluorescent Timer, Fast-FT \cite{Subach2009}. Similar to mCherry, Fast-FT emits blue fluorescence soon after translation. However, the chromophore spontaneously matures into the red form. Our previous experimental measurement of the maturation half-time was 4 hours \cite{Bending2018EMBO}. The two different types of Timer fluorescence—Timer Blue fluorescence from the immature chromophore and Timer Red fluorescence from the mature chromophore—are excited by different lasers (405 nm laser and 561 nm laser, respectively) and analyzed and measured by different detectors (Figure~\ref{fig:fig1}A).

\subsection{Overview of \texttt{TockyPrep}}

\begin{figure}[H]
  \centering
  \includegraphics[width=0.9\textwidth]{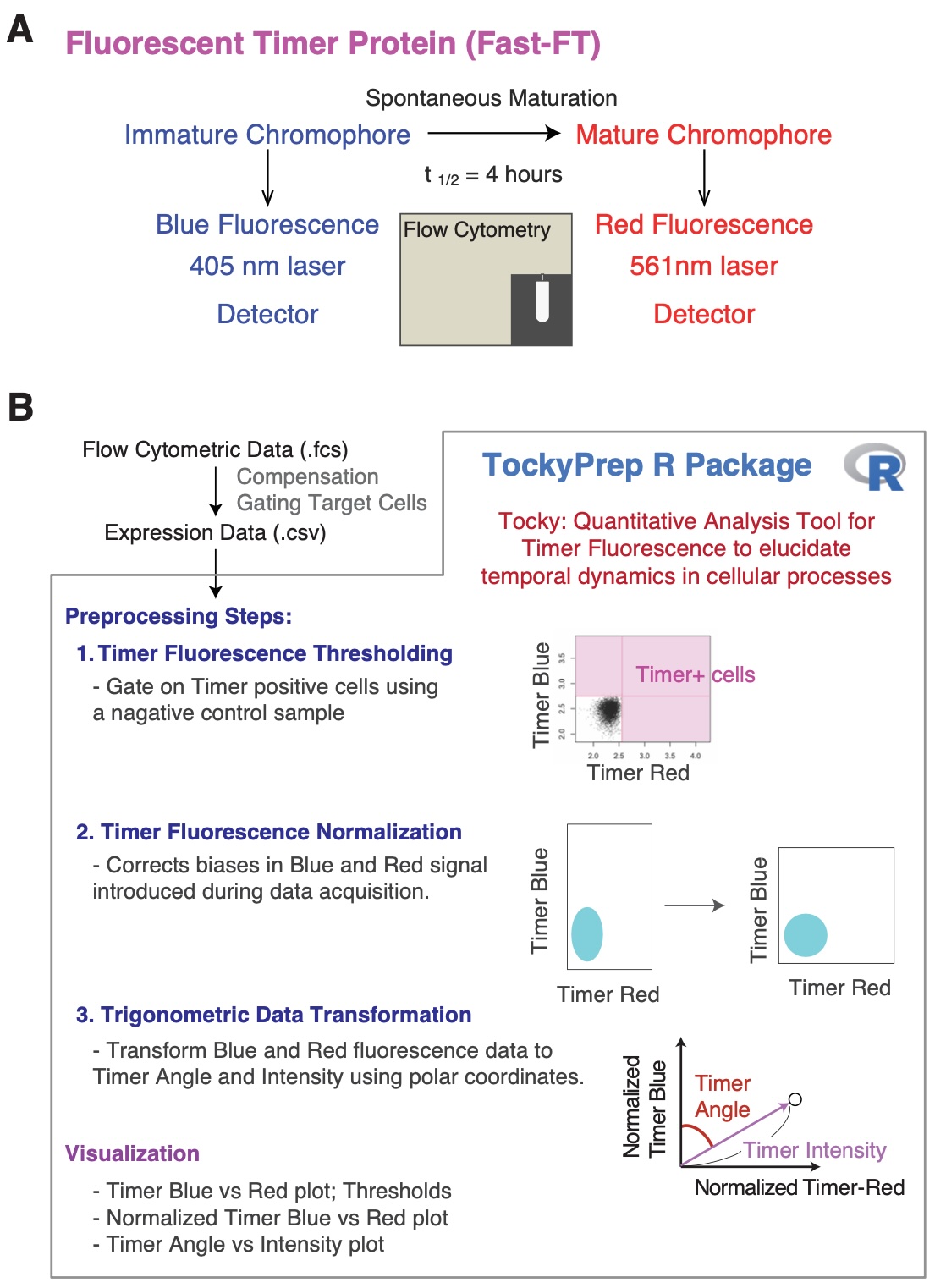}
  \caption{\textbf{Fluorescent Timer protein and workflow for Timer fluorescence data preprocessing using the \texttt{TockyPrep} package.} (A) Diagram illustrating that Timer Blue fluorescence from the immature chromophore is excited by a 405 nm laser, and Timer Red fluorescence from the mature chromophore is excited by a 561 nm laser. These fluorescences are detected by different detectors during flow cytometric analysis. (B) Overview of the \texttt{TockyPrep} workflow, showing data setup using \texttt{prep\_tocky} and preprocessing steps executed by \texttt{timer\_transform}, which include Timer Thresholding, Timer Fluorescence Normalization, and Trigonometric Transformation.}
  \label{fig:fig1}  
\end{figure}

The Tocky technology is an integrated method that combines experimental tools utilizing Fluorescent Timer reporter systems with computational analysis of Timer fluorescence. Using flow cytometric analysis, Tocky is thus a quantitative analysis tool for Timer fluorescence to elucidate temporal dynamics in cellular activities \cite{Bending2018JCB, Ono2024} (Figure~\ref{fig:fig1}A).

The \texttt{TockyPrep} R package offers a structured approach to preprocessing flow cytometric data of Fluorescent Timer proteins, enabling effective analysis of Timer fluorescence and thereby enhancing the Tocky approach. It is built entirely using base R functions, ensuring broad compatibility and ease of use. The package includes two principal functions: \texttt{prep\_tocky} for data setup and \texttt{timer\_transform} for performing the preprocessing steps (Figure~\ref{fig:fig1}B).

To use \texttt{TockyPrep}, flow cytometric data should first be preprocessed using external software for quality control and target population identification. The package accepts expression data as CSV files, each corresponding to a flow cytometric sample. These data are then imported into R for preprocessing, which includes:

\begin{itemize}
    \item \textbf{Timer Thresholding}: Focuses analysis on Timer-positive cells only, avoiding inclusion of autofluorescent cells. A negative control sample is used to define the thresholds for Timer Blue and Red fluorescence.
    \item \textbf{Timer Fluorescence Normalization}: Removes bias between immature (blue) and mature (red) Timer fluorescence introduced during data acquisition.
    \item \textbf{Trigonometric Transformation}: Transforms fluorescence data to calculate Timer Angle and Timer Intensity, representing temporal information and signal strength, respectively.
\end{itemize}

\subsection{Workflow}

\begin{figure}[H]
  \centering
  \includegraphics[width=1\textwidth]{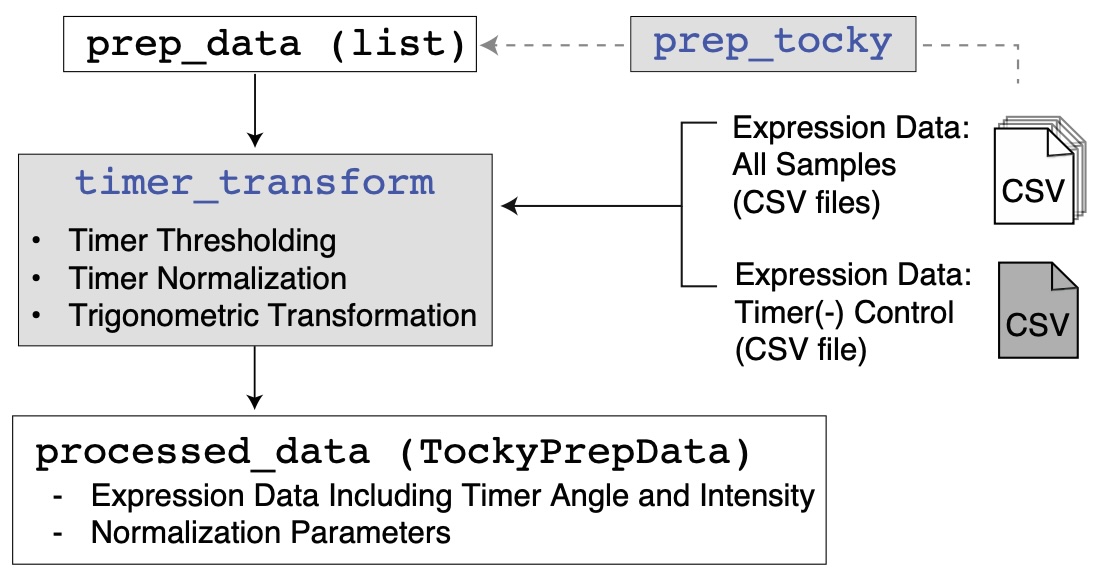}
\caption{\textbf{The two central functions in the \texttt{TockyPrep} implementation, \texttt{prep\_tocky} and \texttt{timer\_transform}.} The function \texttt{prep\_tocky} determines the sample CSV files for processing and identifies key variables, enabling \texttt{timer\_transform} to execute data preprocessing steps in a batch manner.}
  \label{fig:fig2}  
\end{figure}

The \texttt{TockyPrep} package workflow involves two central functions that automate the preprocessing steps (Figure~\ref{fig:fig2}):

\begin{enumerate}
    \item \texttt{prep\_tocky}: Prepares the data by identifying key variables and designating control files.
    \item \texttt{timer\_transform}: Executes the preprocessing steps—Timer Thresholding, Timer Fluorescence Normalization, and Trigonometric Transformation.
\end{enumerate}

\subsection{Data Preparation with \texttt{prep\_tocky}}

The \texttt{prep\_tocky} function provides a convenient preparatory analysis of flow cytometry data files, ensuring a streamlined setup for Timer transformation analysis. It performs the following primary steps:

\begin{itemize}
    \item \textbf{File Selection and Directory Setup}: Defines the path to data files, sets an output directory, and filters for control and sample files based on user input or default settings. The user is prompted to select a negative control file (\texttt{negfile}).
    \item \textbf{Common Variable Identification}: Identifies common variables in the header of each sample file, retaining only variables shared by all files to ensure consistency across datasets.
    \item \textbf{Output Generation}: Generates and exports lists of the selected sample files, control files, and identified channels to the output directory for reference during data transformation. Returns a summary list containing paths to control files, selected sample files, and the standardized variables available for analysis.
\end{itemize}

The \texttt{prep\_tocky} function thus provides an organized and reproducible approach to initializing flow cytometric data for Timer fluorescence analysis, ensuring consistent variable definitions and setting up essential file identities and channels for downstream data preparation.

\subsection{Timer Transformation with \texttt{timer\_transform}}

\begin{figure}[H]
  \centering
  \includegraphics[width=0.7\textwidth]{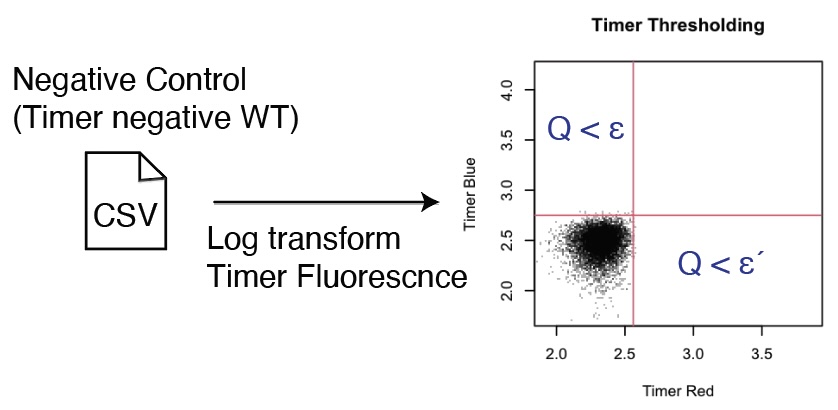}
  \caption{\textbf{Setting thresholds for Timer fluorescence.} The thresholds for Timer Blue and Red fluorescence are customizable. The \texttt{timer\_transform} function allows for automated determination via quantile settings or manual selection through an interactive session utilizing a 2D plot of Timer Blue and Red fluorescence.}
  \label{fig:fig3}  
\end{figure}

The \texttt{timer\_transform} function automates the preprocessing of Timer fluorescence data, performing the following steps:

\begin{itemize}
    \item Accepts the output from the \texttt{prep\_tocky} function, including file paths and variable names.
    \item Sets thresholds for blue and red fluorescence to identify Timer-positive cells, avoiding the inclusion of autofluorescent cells. Threshold values can be determined interactively or through automatic quantile-based methods (Figure~\ref{fig:fig3}).
    \item Normalizes Timer fluorescence data using autofluorescence data to remove bias introduced during data acquisition (Figure~\ref{fig:fig4}).
    \item Applies a trigonometric transformation to calculate Timer Angle and Timer Intensity, representing temporal dynamics and signal strength, respectively.
    \item Produces outputs including transformed data frames with angle and intensity values, normalization parameters, cell counts for each sample, and a sample definition file categorizing each dataset for further analysis.
\end{itemize}

\subsection{Timer Fluorescence Normalization}

\begin{figure}[H]
  \centering
  \includegraphics[width=1\textwidth]{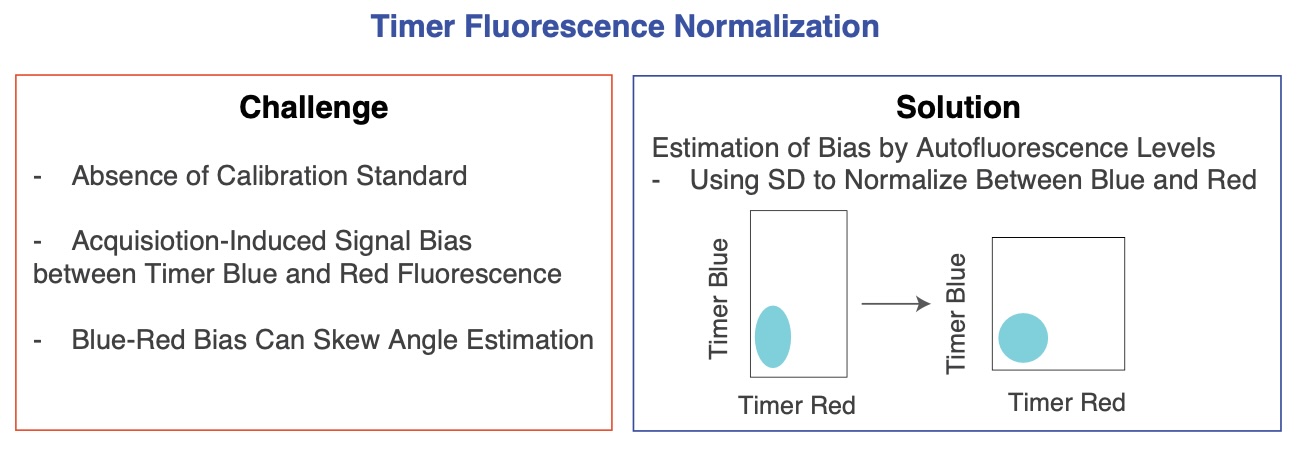}
  \caption{\textbf{Timer Fluorescence Normalization.} This figure illustrates the challenges in Timer fluorescence data analysis: (1) the absence of a calibration standard for Timer Blue, (2) acquisition settings that induce signal bias between Timer Blue and Red fluorescence, and (3) the resultant skew in Timer Angle estimation due to these biases. To address these issues, the \texttt{TockyPrep} package includes a function for Timer fluorescence normalization. This function utilizes autofluorescence levels to estimate inherent signal biases and employs MAD to normalize differences between Blue and Red fluorescence, thereby enhancing the accuracy of Timer Angle calculations.}
  \label{fig:fig4}  
\end{figure}

Timer fluorescence data analysis currently faces several significant challenges (Figure~\ref{fig:fig4}). Firstly, there is a notable absence of a calibration standard for Timer Blue and Timer Red, which complicates accurate measurements. This means that there is no available method to robustly determine the absolute expression levels of the two types of Timer fluorescence. Secondly, the settings used during data acquisition can induce biases between Timer Blue and Red fluorescence. Thirdly, these biases often result in a skew in Timer Angle estimations, which can misrepresent true biological phenomena.

To address these challenges, the \texttt{TockyPrep} package introduces a robust function specifically designed for Timer fluorescence normalization. This function leverages autofluorescence levels to accurately estimate inherent signal biases. Moreover, it applies MAD by default as the normalization factor. These adjustments correct disparities between Timer Blue and Red fluorescence signals, thereby enhancing the precision of Timer Angle calculations.

 \newpage

\section{Results}

The data preprocessing using the \texttt{TockyPrep} package is highly effective and can be completed promptly. For example, the \texttt{timer\_transform} function using the Nr4a3 Tocky dataset comprising 33 flow cytometry data files totaling 30 MB was completed within 3 seconds.

Since the Timer Angle is calculated based on the fluorescence intensities of Timer Blue and Red fluorescence in individual cells (Figure~\ref{fig:fig1}), an imbalance of data intensities between Timer Blue and Red fluorescence can result in severely biased Timer Angle calculations. Accordingly, this section focuses on the impact of Timer fluorescence normalization.

In this investigation, we analyzed the flow cytometric dataset that analyzed activated T cells obtained from Nr4a3-Tocky::OT-II double transgenic mice\cite{Bending2018JCB}. These T cells, which express the transgenic TCR OT-II, were activated with 1~\(\mu\)M Ova\textsubscript{(323-339)} peptide to assess the dynamic transcriptional activity of Nr4a3 following antigen stimulation in a time course analysis \cite{Bending2018JCB}. This dataset offers opportunities to address how Tocky data preprocessing is applied.

\subsection{Impact of Timer Fluorescence Normalization on Timer Angle Calculation}

\begin{figure}[H]
  \centering
  \includegraphics[width=1\textwidth]{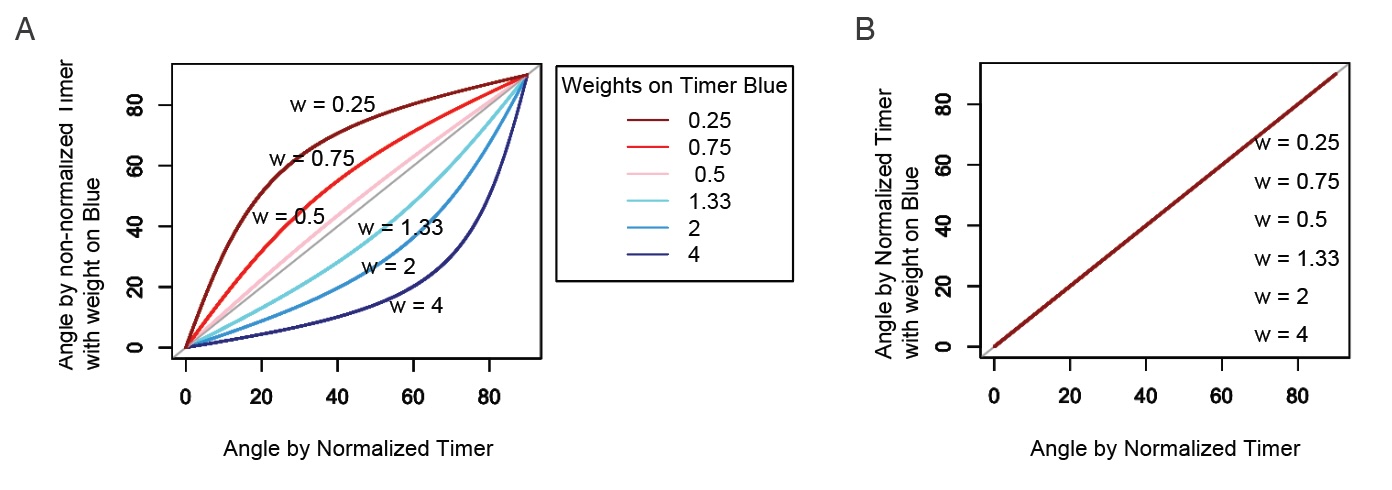}
  \caption{\textbf{Impact of Timer Fluorescence Normalization on Timer Angle Calculation.} (A) 2D plot showing the relationship between normalized and non-normalized Timer Angles with varying weights applied to Timer Blue fluorescence. (B) Effect of applying normalization to the simulated data shown in (A), demonstrating the correction of Timer Angle estimates.}
  \label{fig:fig5}  
\end{figure}

To assess the effect of biased data acquisition on Timer Angle calculations, we manipulated the Nr4a3 Tocky dataset by applying various multipliers to the Timer Blue fluorescence while keeping Timer Red fluorescence unchanged. This approach simulated different levels of signal intensity bias in Timer Blue fluorescence to observe its impact on calculated Timer Angles. As Figure~\ref{fig:fig5} demonstrates, increasing the Timer Blue fluorescence led to a decrease in the estimated Timer Angle, indicating a distortion in the angle calculation due to the artificial enhancement of blue signal intensity.

Conversely, normalizing the Timer Blue and Red fluorescence effectively eliminated these distortions. The normalization process adjusted the calculated angles to remove the bias introduced by varying the intensity of the Timer Blue fluorescence.

\subsection{Analyzing Activated T Cells from Nr4a3 Tocky Mice Using \texttt{TockyPrep}}

\begin{figure}[H]
  \centering
  \includegraphics[width=1\textwidth]{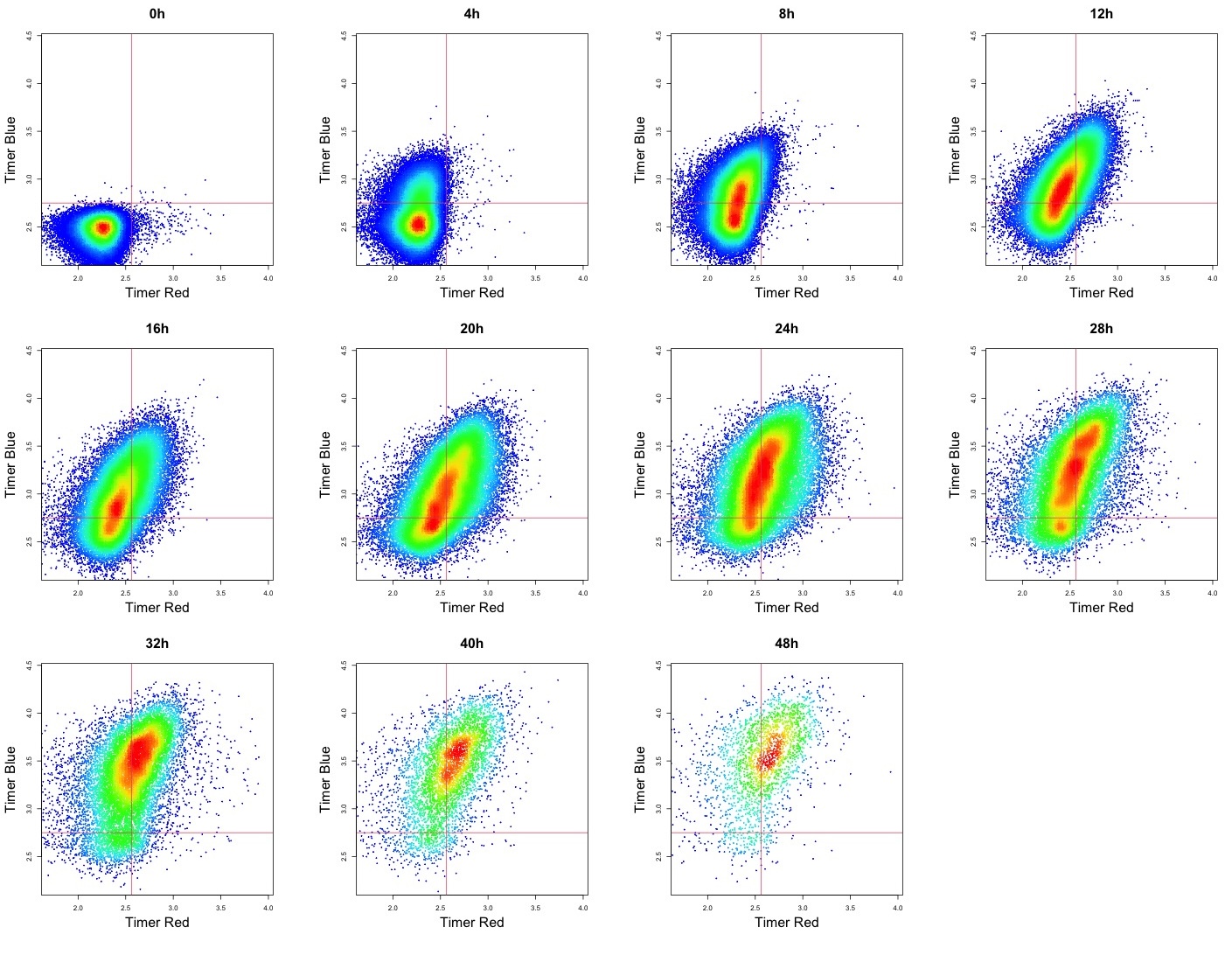}
\caption{\textbf{Visualization of Timer Thresholding Using \texttt{plot\_tocky}.} This figure shows Timer Blue and Red fluorescence of T cells before normalization. T cells from Nr4a3-Tocky::OT-II double transgenic mice were activated by Ova peptide and analyzed at the indicated time points \cite{Bending2018JCB}. Quadrant gates illustrate the thresholds for Timer Blue and Red fluorescence set by \texttt{timer\_transform}.}
  \label{fig:fig6}  
\end{figure}

The \texttt{TockyPrep} package includes a key plot function, \texttt{plot\_tocky}, which visualizes Timer fluorescence variables to understand the effects of thresholding, normalization, and transformation. Using the option \texttt{plot\_mode = "Timer fluorescence"}, \texttt{plot\_tocky} shows raw Timer fluorescence, demonstrating how Timer thresholds are applied to experimental samples (Figure~\ref{fig:fig6}). In Figure~\ref{fig:fig6}, dynamic changes of Timer profiles upon TCR stimulation are thus visualized in conventional 2D plots for Timer fluorescence: Timer Blue fluorescence is quickly induced shortly after activation, followed by an increase in Timer Red fluorescence. Importantly, the quadrant gates displayed in the output from \texttt{plot\_tocky} represent the Timer thresholds that have been set by \texttt{timer\_transform}, assisting the quality check of Timer data preprocessing.

Notably, T cells at the 0-hour time point include Timer Blue-Red\(^{+}\) cells, comprised of self-reactive T cells that accumulate Timer Red fluorescence in vivo \cite{OnoSatou2024}. Since this accumulation occurs during antigen recognition in vivo and not from in vitro antigen stimulation, this time point is excluded from downstream analysis.

\begin{figure}[H]
  \centering
  \includegraphics[width=1\textwidth]{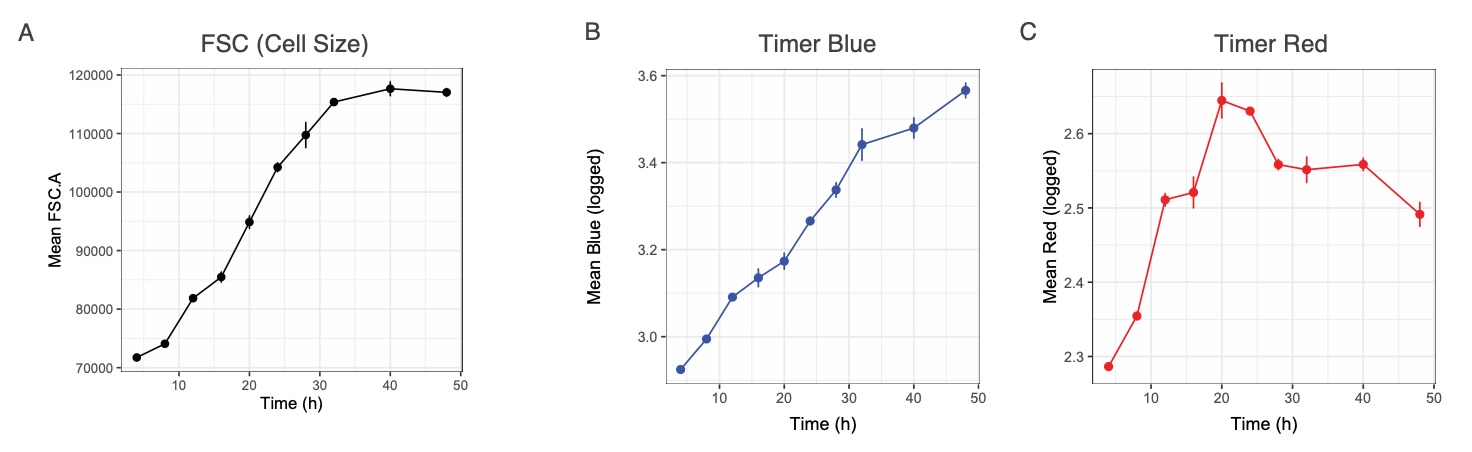}
  \caption{\textbf{Changes in FSC and Timer Fluorescence After T-Cell Activation.} (A) FSC measurements showing increased cell size after activation. (B) Timer Blue fluorescence increases after activation. (C) Timer Red fluorescence increases after activation. Note that the time point zero is excluded due to the accumulation of self-reactive T cells that accumulate Timer red fluorescence.}
  \label{fig:fig7}  
\end{figure}

Once T cells recognize an antigen, they receive TCR signaling and become activated, exhibiting increased cell size as measured by forward scatter (FSC) and elevated Timer expression. As expected, FSC significantly increases following activation (Figure~\ref{fig:fig7}A). Similarly, both Timer Blue and Red fluorescence intensify post-activation, as demonstrated by the analysis of mean fluorescence intensities (Figure~\ref{fig:fig7}B and C). 

\begin{figure}[H]
  \centering
  \includegraphics[width=1\textwidth]{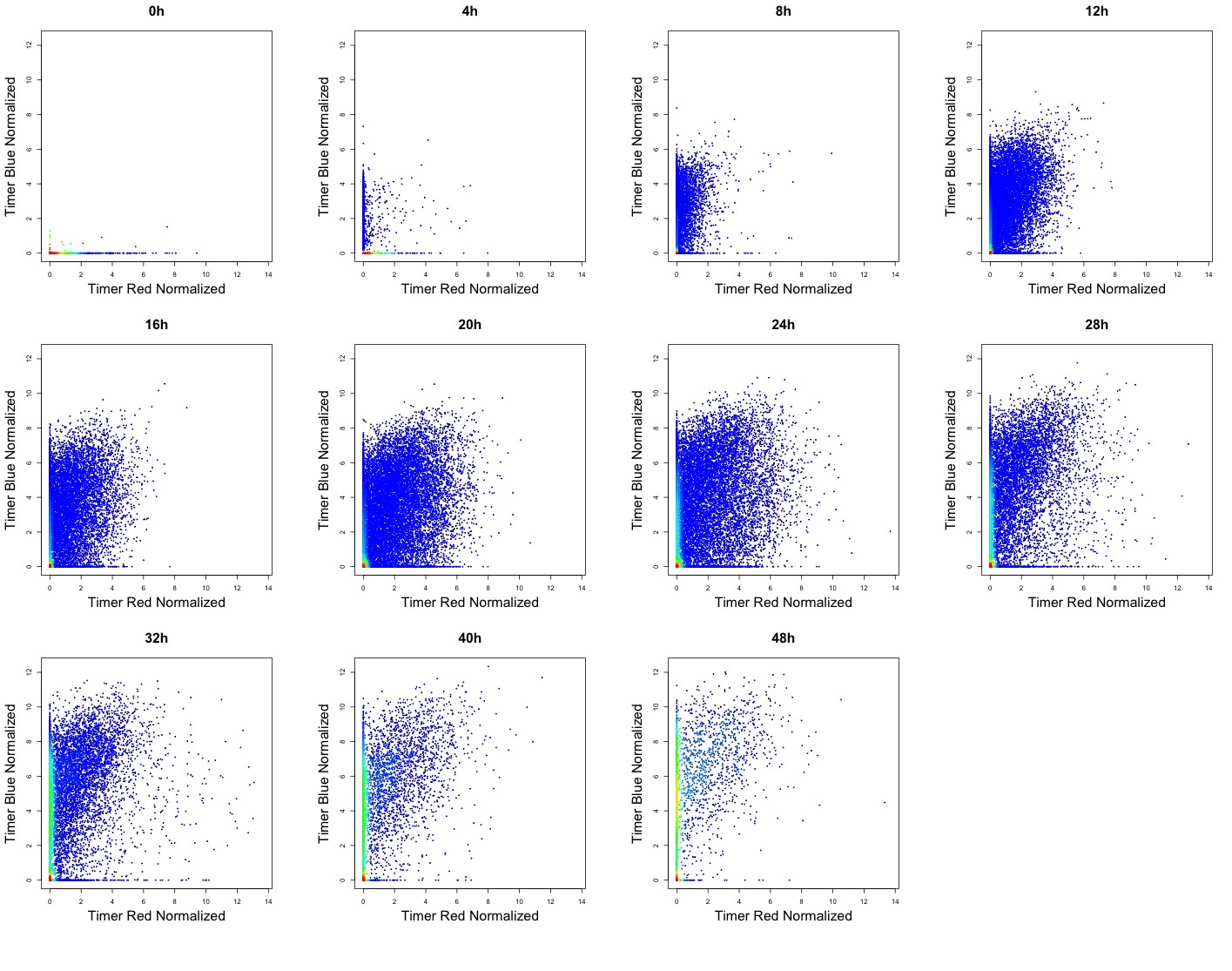}
\caption{\textbf{Visualization of Normalized Timer Fluorescence Using \texttt{plot\_tocky}.} This figure displays the normalized Timer Blue and Red fluorescence, visualized using \texttt{plot\_tocky}. The data shown here are the same as those in Figure 7, but presented with the option \texttt{plot\_mode = "Normalized Timer fluorescence"}.}
  \label{fig:fig8}  
\end{figure}

After applying the \texttt{timer\_transform} function, we utilized \texttt{plot\_tocky} with the option \texttt{plot\_mode = "Normalized Timer fluorescence"} to visualize the effects of Timer thresholding and normalization on Timer Blue and Red fluorescence (Figure~\ref{fig:fig8}). Importantly, the data displayed in this visualization reflect not only the normalization of Timer Blue and Red fluorescence but also the application of Timer thresholding, ensuring that all analyzed cells are Timer-positive.

\begin{figure}[H]
  \centering
  \includegraphics[width=1\textwidth]{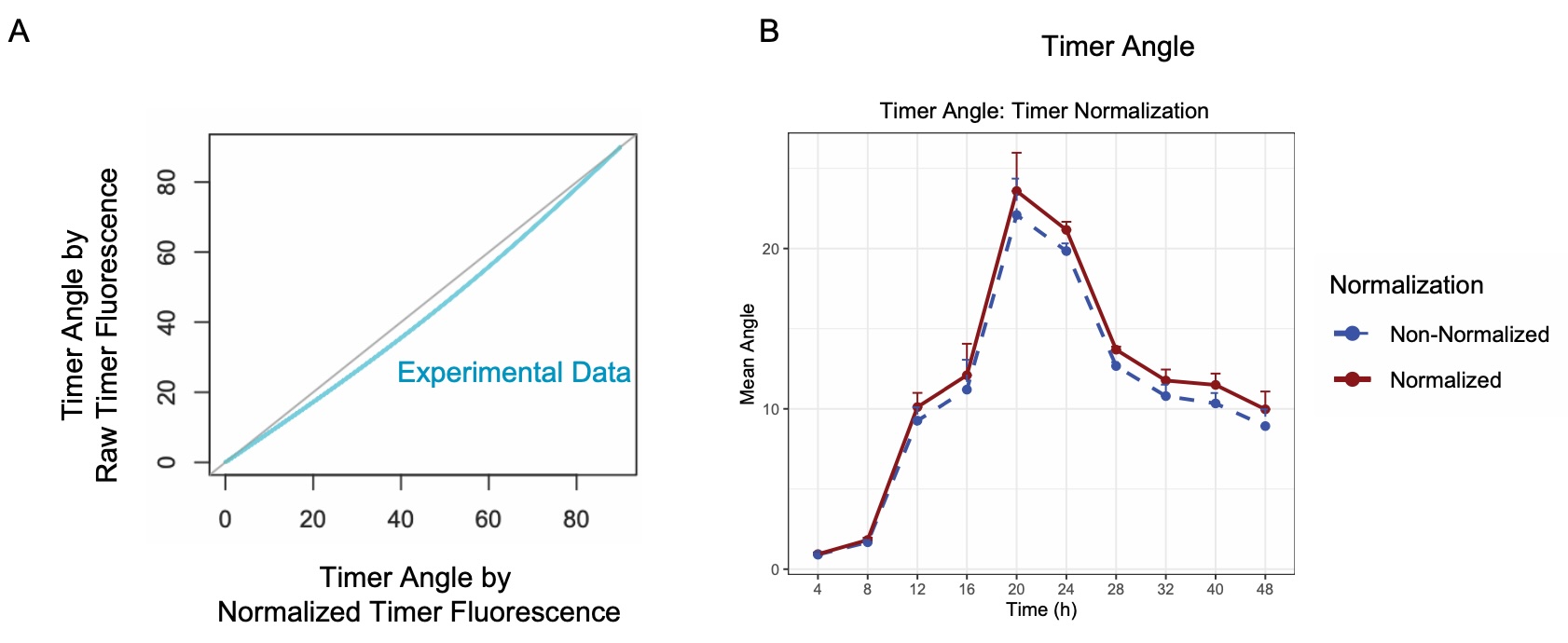}
  \caption{\textbf{Impact of Timer Normalization on Timer Angle Calculation Using Experimental Data.} The figure illustrates that normalization results in minor adjustments to Timer Angle estimations, correcting slight biases present in non-normalized data.}
  \label{fig:fig9}  
\end{figure}

In the experimental dataset, the effects of Timer fluorescence normalization were modest (Figure~\ref{fig:fig8}). This is likely due to the near-optimal settings of the instrumentation. The estimation of Timer Angle using both normalized and non-normalized Timer fluorescence data exhibited only slight differences, with the non-normalized data showing marginally lower estimations (Figure~\ref{fig:fig9}).

\begin{figure}[H]
  \centering
  \includegraphics[width=1\textwidth]{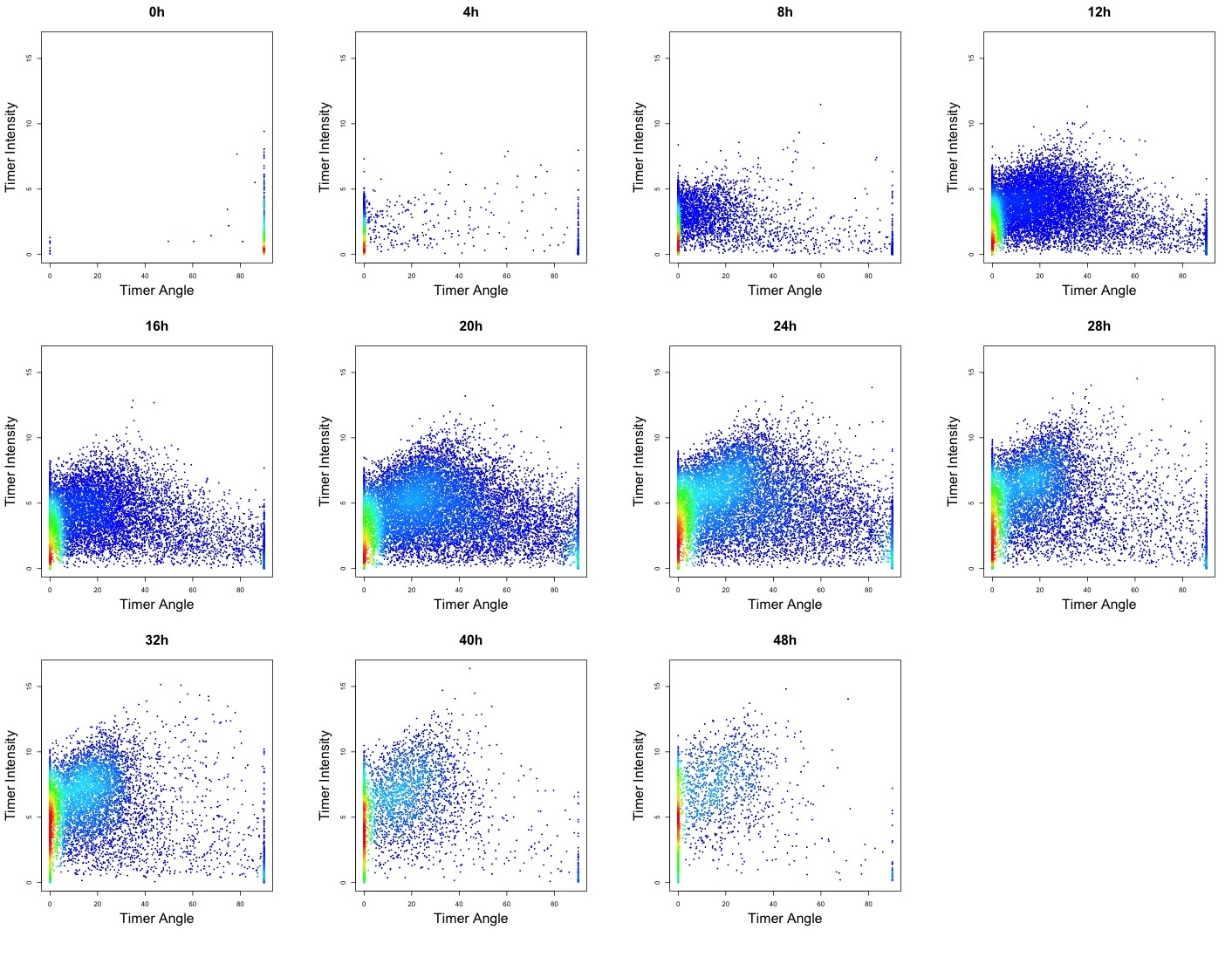}
\caption{\textbf{Visualization of Timer Angle and Intensity Using \texttt{plot\_tocky}.} Two-dimensional plots showing Timer Angle and Intensity were generated using the \texttt{plot\_tocky} function with the option \texttt{plot\_mode = "Timer Angle and Intensity."}. These plots display data that has been normalized and transformed into polar coordinates.}  \label{fig:fig10}  
\end{figure}

Finally, we examined Timer Angle and Intensity using the function \texttt{plot\_tocky} with the option \texttt{plot\_mode = "Timer Angle and Intensity."}, which produces group-wise 2D plots for Timer Angle and Intensity. The Timer Angle-Intensity plots effectively show the progressive increase of Timer expression from low Timer Angle values, gradually accumulating cells with high Timer Angle and Intensity values (Figure~\ref{fig:fig10}).

\subsection{Exploring Timer Data Transformation Parameters}
\begin{figure}[H]
  \centering
  \includegraphics[width=1\textwidth]{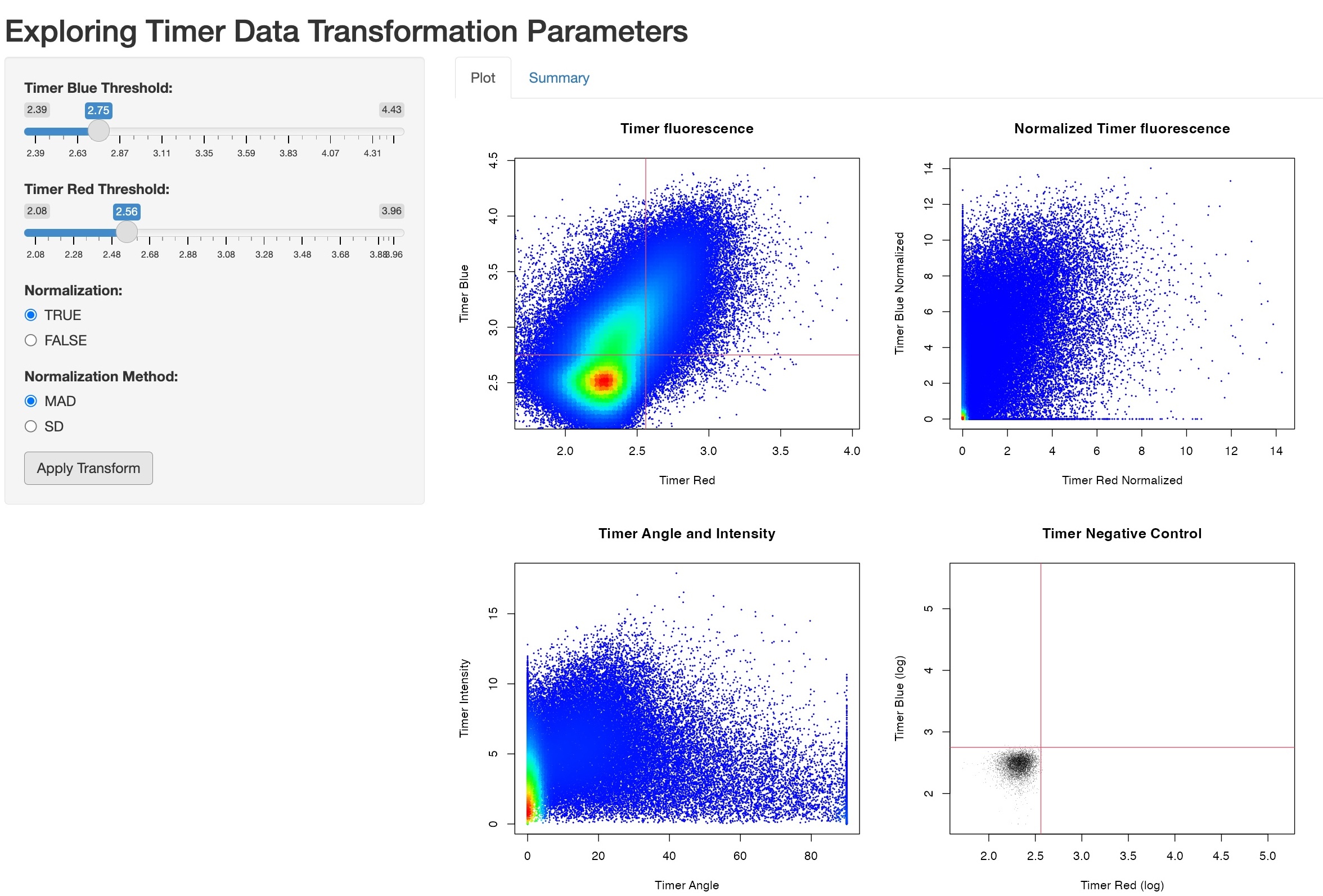}
\caption{\textbf{Exploration of Timer Data Transformation Parameters Using \texttt{explore\_timer\_transform}.} This figure displays a screenshot of the interactive application outputs. The sidebars enable users to set thresholds for Timer Blue and Timer Red fluorescence and to choose a normalization method. The plot panels include 2D plots for raw Timer fluorescence (Upper Left), normalized Timer fluorescence (Upper Right), and Timer Angle and Timer Intensity (Lower Left) using all concatenated expression data. Additionally, a 2D plot of negative control data for raw Timer fluorescence is shown, with Timer thresholds determined by the sidebar settings.}  \label{fig:fig11}

\end{figure}

By using the function \texttt{explore\_timer\_transform}, the parameter space can be explored by adjusting parameters for Timer fluorescence normalization and transformation. Executing \texttt{explore\_timer\_transform} launches an interactive HTML application, enabling users to experiment with changing Timer fluorescence thresholds and selecting different normalization methods on their flow cytometric Timer data (Figure~\ref{fig:fig11}).

\section{Discussion}

The \texttt{TockyPrep} package offers essential data preprocessing methods for the analysis of Timer fluorescence using flow cytometric data. To our knowledge, this represents the first effort to develop quantitative analysis tools specifically for Timer fluorescence data analysis. 

Our investigations identified the normalization of Timer fluorescence between Timer blue and red fluorescence was key for quantitative analysis of Timer maturation using the trigonometric transformation. Since the current method relies on autofluorescence data, ubiquitous adjustments through normalization using the MADs or SDs of Timer Blue and Red fluorescence may not achieve the most optimized results for cells with extremely high expression. Still, our current investigations demonstrated the importance of Timer fluorescence normalization and the utility of the current method and implementation, providing reasonable and feasible solutions. Future studies can focus on the development of experimental standards to further enhance the normalization methods.

Our previous study utilized FSC data to correct size-dependent effects in Timer fluorescence \cite{Bending2018JCB}, a method not implemented in the current development. Since traditional flow cytometry often shows higher background in the blue fluorescence channel, including Timer Blue fluorescence, the principle underlying FSC correction remains valid. However, in analyzing real datasets, we identified two significant issues. First, the FSC correction method depends on Timer-negative control samples, which typically lack activated cells with larger sizes. These cells are often removed during the quality control stage, complicating the determination of optimal FSC correction parameters across the full range of FSC and Timer fluorescence. Second, modern flow cytometers, especially spectral analyzers, show greater robustness against cell size effects. Consequently, we opted not to implement the FSC correction method in the current version of the package.

To date, various Fluorescent Timer protein reporter systems have been reported. The initial murine reporter systems utilizing Fast-FT were Nr4a3-Tocky, designed to analyze the TCR signal downstream gene Nr4a3, and Foxp3-Tocky, developed for the regulatory T cell-specific gene Foxp3 \cite{Bending2018JCB, Bending2018EMBO}. These systems pioneered the use of Fast-FT in analyzing transcriptional and cellular dynamics across cellular biology and immunological studies \cite{Ono2024}. Subsequent developments have expanded to include HIV-Tocky for the long terminal repeat (LTR) of HIV \cite{Reda2024}, Nur77-Tempo for analyzing TCR signal downstream Nur77 (Nr4a1), another member of the Nr4a family, similar to Nr4a3-Tocky \cite{Tempo2022}, and H2B-FT for a histone gene \cite{Eastman2020}. In addition, other fluorescent Timer proteins like DsRed-E5 have been utilized to develop a transgenic reporter for the proglucagon gene Gcg \cite{Himuro2024}. Notably, Nr4a3-Tocky has been the most frequently used for the broadest range of applications including tumour immunology and immunotherapy \cite{Hassan2022}\cite{Bozhanova2022} NKT cell biology\cite{Bortoluzzi2021}, and autoimmune disease\cite{Bending2018JCB}.

Despite these advances, until now, there has been no tool developed specifically to quantitatively analyze flow cytometric Timer fluorescence data. We anticipate that the \texttt{TockyPrep} tool will aid in the analysis of Timer fluorescence data from these and other reporter systems.

Future work for Tocky analysis will include downstream data analysis which fully harnesses the power of the Timer system. To achieve the development of quantitative methods and visualization tools using Timer fluorescence data, the current study and the \texttt{TockyPrep} package provide an important basis and preparation for future development.

\section{Conclusion}

In this study, we established data preprocessing methods dedicated to Timer fluorescence analysis. These methods can improve the quantitative analysis of Timer fluorescence data using the trigonometric transformation of Timer fluorescence data. The current study identified the importance of Timer fluorescence normalization and related data preprocessing methods. These have been implemented in the R package \texttt{TockyPrep}.

\section{Abbreviations}

\begin{itemize}
\item Tocky: Timer-of-cell-kinetics-and-activity
\item FSC: Forward Scatter
\item TCR: T-cell Receptor
\item MAD: Median Absolute Deviation
\item SD: Standard Deviation
\end{itemize}

\section{Declarations}

\begin{itemize}
\item Ethics approval and consent to participate: Not applicable.

\item Consent for publication: Not applicable.

\item Availability of data and materials: The \texttt{TockyPrep} R package is freely available through the MonoTockyLab GitHub repository (\url{https://github.com/MonoTockyLab/TockyPrep}). A comprehensive vignette and manual pages are accessible at \url{https://monotockylab.github.io/TockyPrep/}. The flow cytometric dataset presented in this study is available as a part of the \texttt{TockyPrep} package.

\item Competing interests: The author has filed patents for methodologies that are utilised for the computational analysis of Fluorescent Timer data. These methodologies are distinct and separate from the algorithms reported in this manuscript.

\item Funding: This work was supported by a Cancer Research UK (CRUK) Programme Foundation Award (grant number DCRPGF-100007), a Medical Research Council (MRC) grant (grant number MR/S000208/1), and a Biotechnology and Biological Sciences Research Council (BBSRC) David Phillips Fellowship (grant number BB/J013951/2).

\item Authors' contributions: MO conceived the study, wrote the code, performed data analysis, and wrote the manuscript.

\end{itemize}

\section{Availability and requirements}

\begin{itemize}
\item Project name: \texttt{TockyPrep}
\item Project home page: \url{https://github.com/MonoTockyLab/TockyPrep}
\item Project documentation: \url{https://MonoTockyLab.github.io/TockyPrep}
\item Operating system(s): Linux, Mac OS, Windows
\item Programming language: R
\item Other requirements: R 4.2.0 or higher
\item License: Apache 2.0
\item Any restrictions to use by non-academics: none
\end{itemize}

\bibliographystyle{unsrt}
\bibliography{TockyPrep}

\end{document}